\documentclass{INTERSPEECH2023}
\usepackage{xcolor}
\usepackage{arydshln}
\usepackage {booktabs}
\usepackage {subcaption}
\definecolor{EMgray}{gray}{0.45}

\interspeechcameraready 

\title{Masked Modeling Duo for Speech: Specializing General-Purpose Audio Representation to Speech using Denoising Distillation}
\name{Daisuke Niizumi, Daiki Takeuchi, Yasunori Ohishi, Noboru Harada, and Kunio Kashino}

\address{NTT Corporation, Japan}
\email{daisuke.niizumi@ntt.com}

\begin{document}

\maketitle

\begin{abstract}
Self-supervised learning general-purpose audio representations have demonstrated high performance in a variety of tasks. Although they can be optimized for application by fine-tuning, even higher performance can be expected if they can be specialized to pre-train for an application.
This paper explores the challenges and solutions in specializing general-purpose audio representations for a specific application using speech, a highly demanding field, as an example.
We enhance Masked Modeling Duo (M2D), a general-purpose model, to close the performance gap with state-of-the-art (SOTA) speech models. To do so, we propose a new task, denoising distillation, to learn from fine-grained clustered features, and M2D for Speech (M2D-S), which jointly learns the denoising distillation task and M2D masked prediction task.
Experimental results show that M2D-S performs comparably to or outperforms SOTA speech models on the SUPERB benchmark, demonstrating that M2D can specialize in a demanding field.
\end{abstract}
\noindent\textbf{Index Terms}: speech representation learning, general-purpose audio representation, denoising, distillation, specialization

\section{Introduction}
The self-supervised learning general-purpose audio representations (generic models), pre-trained on a large-scale audio dataset, have shown promising performance on a variety of environmental, musical, and speech tasks.
While the generic models have demonstrated versatile performance \cite{gong2022ssast,Baade2022MAE-AST,niizumi2022msm-mae,niizumi2022M2D,chen2022beats}, they have not fully demonstrated their usefulness in focused applications such as automatic speech recognition, where SOTA performance is required.

On the other hand, highly advanced speech models have been investigated to meet intense demand. In addition to effective self-supervised learning using contrastive loss and masked prediction \cite{LIU2022ASurvey,baevski2020wav2vec2,Hsu2021HuBERT,Chen2022WavLM}, SOTA models make use of the learning of discrete representations by quantization \cite{baevski2020wav2vec2}, pseudo-label generation by clustering \cite{Hsu2021HuBERT,Chen2022WavLM}, and denoising for learning robust representation \cite{Chen2022WavLM}.


We believe generic models can provide even higher performance when pre-trained specifically for a particular application rather than just fine-tuning for the application. This paper explores the challenges and possibilities of specialization using a generic model, masked modeling duo (M2D) \cite{niizumi2022M2D}, in speech as an example.
The question we aim to answer is: \textit{Can a general-purpose audio representation be useful in a competitive field?}

To specialize a generic model in speech, we found that the challenges lie in incorporating speech-specific techniques.
While a generic model can easily switch the pre-training dataset to a speech corpus, we found it is not yet powerful enough to compete with SOTA speech models. Therefore, resorting to field-specific extensions (e.g., learning from clustered features) is inevitable.


With a necessary extension, we propose M2D for Speech (M2D-S), which extends M2D with a new task, denoising distillation, to learn from fine-grained clustered features. M2D-S also adapts the pre-training dataset and patch size without changing M2D.
Experiments using the SUPERB benchmark \cite{yang2021superb} show that M2D-S performs comparably to or better than SOTA speech models, demonstrating that a generic model can serve as a pre-training framework for a challenging field.

Our contributions are i) proposal of a denoising distillation task for speech representation learning, ii) proposal of M2D-S to specialize a generic model to speech, and iii) demonstration of the extensibility of a generic model by achieving SOTA performance in speech.
Our code is available online\footnote{\scriptsize{\url{https://github.com/nttcslab/m2d/tree/master/speech}}}.

\begin{figure}[t]
  \centering
  \includegraphics[width=1.0\columnwidth]{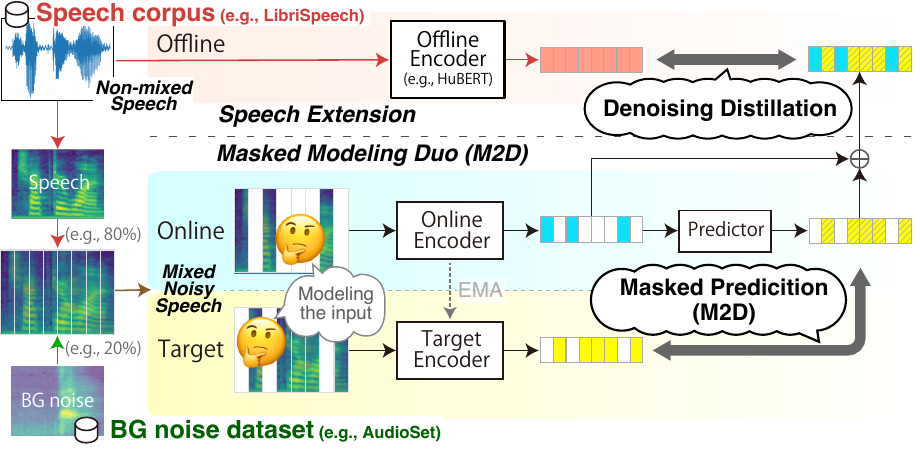}
  \vspace{-10pt}
  \caption{M2D-S extends M2D to speech by adding an offline network for the denoising distillation task. While the online and target networks in M2D learn the masked prediction of the features from noisy speech, the offline network provides pure speech features for learning the denoising distillation.}
  \label{fig:scenario}
  \vspace{-10pt}
\end{figure}

\section{Related Work}\label{sec:related-work}
Self-supervised learning methods that train transformers with masked prediction have shown promising performance in various domains.
Speech representation models, such as Mockingjay \cite{Liu2020Mockingjay} and TERA \cite{Liu2021TERA}, take spectrograms as input. TERA employs a masking strategy for splitting both frequency bins and time steps, similar to generic models.
SOTA models, such as wav2vec2.0 \cite{baevski2020wav2vec2}, BigSSL \cite{zhang2022bigssl}, data2vec \cite{baevski2022data2vec}, HuBERT \cite{Hsu2021HuBERT}, and WavLM \cite{Chen2022WavLM}, typically learn an acoustic feature extraction from the input speech waveform in addition to the representation learning. Notably, models such as wav2vec2.0, HuBERT, and WavLM effectively learn by using discretized pseudo-labels with vector quantization or clustering of pre-trained model features. In addition, WavLM has improved performance on non-ASR tasks through additional masked speech denoising.


Generic models, such as AST \cite{gong2021ast}, SSAST \cite{gong2022ssast}, ATST \cite{Li2022ATST}, MAE-AST \cite{Baade2022MAE-AST}, MSM-MAE \cite{niizumi2022msm-mae}, AudioMAE \cite{huang2022amae}, M2D \cite{niizumi2022M2D}, and BEATs \cite{chen2022beats}, have shown SOTA performance.
They typically take spectrograms as acoustic feature input, split input in both the time and frequency axes, and train a Vision Transformer (ViT) \cite{ViT} while avoiding application-specific techniques such as masking inputs for consecutive time steps, as in speech.

In previous works similar to this paper, Melms et al. \cite{le2023lungsound} and BYOL-S \cite{elbanna22BYOL-S} have specialized a generic model BYOL-A \cite{niizumi2023byol-a} in medical and speech applications.
SSAST has adapted patch size and pre-training dataset and compared it with speech models on SUPERB.
In the NLP domain, LIBERT \cite{lauscher-etal-2020-specializing} specializes BERT \cite{bert} using an additional task for pre-training a lexically-informed BERT, a similar multi-task learning setting to ours.
Previous works that 
created pseudo-labels or distill models include BEATs, HuBERT, WavLM, DistilHuBERT \cite{DistilHuBERT}, MT4SSL \cite{ma2022mt4ssl}, and RobustDistiller \cite{guimares2023robustdistiller}. In particular, concurrent works MT4SSL and RobustDistiller learn multi-tasks, similar to ours, and RobustDistiller auxiliary learns to denoise speech in addition to distillation.

Unlike these previous studies, we attempt to achieve the performance of SOTA speech models by specializing M2D.

\section{Method}\label{sec:method}
M2D-S specializes in speech by adding a speech extension network and its task to M2D without changing M2D.
Table \ref{tab:diff-ssl} lists the major differences between M2D and SOTA speech models.
The key challenge of speech specialization is to implement in M2D-S the features that critically impact speech task performance.

\begin{table}[htb!]
\vspace{-8pt}
\caption{Key differences between models.}
\vspace{-5pt}
\label{tab:diff-ssl}
\centering
\resizebox{\columnwidth}{!}{%
\begin{tabular}{clll}
\toprule
 &  & Speech model & Generic model \\
 & Design choices & (e.g., HuBERT, WavLM) & (M2D) \\
\midrule
(a) & {Pre-training dataset} & Speech (e.g., LibriSpeech & General audio \\
& &  \cite{Panayotov2015LibrispeechAA}) \& noises (WavLM) & (e.g., AudioSet \cite{gemmeke2017audioset}) \\
(b) & Input data format & Raw waveform & Spectrogram \\
(c) & {Input data split} & Split in time steps & Split in both freq./time \\
(d) & Feature extractor & CNN & (not used) \\
(e) & {Discretized} & Clustering pre-trained & (not used) \\
& training signals & model features & \\
(f) & Extra pre-training task & Denoising (WavLM) & (not used) \\
(g) & Masking strategy & Consequent time steps & Random \\
\bottomrule\\
\end{tabular}
}
\vspace{-15pt}
\end{table}

Based on the preliminary experimental results, we focus on (a) using a speech corpus as the dataset, (c) splitting the input only along the time steps, (e) using clustered pre-trained model features as a training signal, and (f) using a denoising task.

In addition, we propose a new task, denoising distillation, which maximizes the use of fine-grained clusters in the distribution of the pre-trained speech model features and performs (e) and (f) simultaneously.

\subsection{Background: Masked Modeling Duo}\label{sec:m2d}
\begin{figure}[tbp]
  \vspace{-10pt}
  \centering
  \includegraphics[width=1.0\columnwidth]{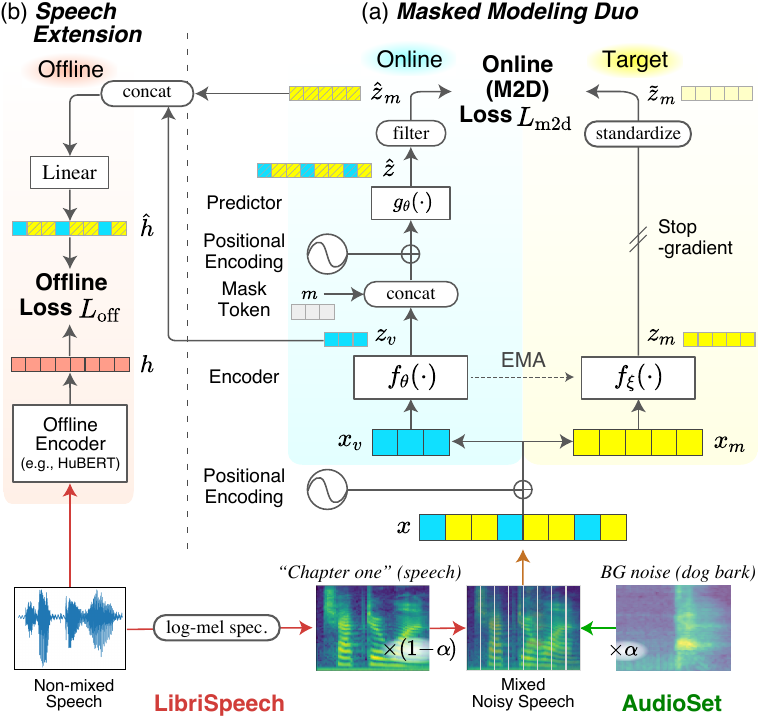}
  \vspace{-15pt}
  \caption{Overview of the M2D-S framework.}
  \label{fig:system}
  \vspace{-15pt}
\end{figure}

M2D is a self-supervised learning framework applicable to 2D structured data input such as images and audio spectrograms, and trains ViT with masked prediction.
As shown in Fig. \ref{fig:system}(a), it consists of two networks, the online and the target, and learns to predict the target output representations using the online output representations.
Unlike speech models (e.g., HuBERT), M2D takes a spectrogram (e.g., 80 frequency bins and 208 time steps) as input, which is split into patches (e.g., $16\times 16$) and treated as a series (e.g., $(80/16)\times(208/16)=65$ patches).

M2D splits the input data $x$ into patches, adds positional encoding, and randomly selects a number of patches according to a masking ratio as masked patches $x_m$ (e.g., 60\% of the input) and the rest as visible patches $x_v$ (e.g., the remaining 40\%).

The online network with a set of weights $\theta$ encodes $x_v$ using the online encoder $f_\theta$ into the representation $z_v = f_\theta(x_v)$.
It concatenates the learnable masked tokens $m$ to $z_v$, adds the position encoding $p$, and inputs to the predictor $g_\theta$ to predict the representation $\hat{z} = g_\theta(\text{concat}(z_v, m) + p)$.
It then outputs the prediction result $\hat{z}_m = \{\, \hat{z}[i] \mid i \in I_M \,\}$ of the masked patch representations, where $I_M$ is the set of masked patch indices.

The target network defined by parameter $\xi$ outputs the representation $z_m = f_\xi(x_m)$ and standardizes it to the final target output $\tilde{z}_m = ({z_m - \text{mean}{(z_m)}})/{\sqrt{\text{var}{(z_m)}}}$.

The loss is calculated using the online prediction $\hat{z}_m$ against the target output $\tilde{z}_m$ as a training signal by the mean square error (MSE) of $l_2$-normalized $\hat{z}_m$ and $\tilde{z}_m$:
\vspace{-0.2cm}
\begin{equation}
L_\text{m2d} \triangleq ||l_2(\hat{z}_m) - l_2(\tilde{z}_m)||^2_2 = 2 - 2 \cdot \frac{\langle \hat{z}_m, \tilde{z}_m \rangle }{||\hat{z}_m||_2 \cdot ||\tilde{z}_m||_2},
\label{eq:eq-byol-mse}
\vspace{-0.2cm}
\end{equation}
where $\langle\cdot, \cdot\rangle$ denotes the inner product.

The M2D framework updates parameters $\theta$ only by minimizing the loss $L_\text{m2d}$ as depicted by the stop-gradient in Fig. \ref{fig:system} (a), and updates $\xi \leftarrow \tau \xi + (1 - \tau) \theta$ as an exponential moving average of $\theta$ with a decay rate $\tau$.

M2D exploits the momentum encoder to learn effective representations from the target network.
After the training, only the $f_\theta$ is used as a pre-trained model in downstream tasks.

\subsection{Denoising Distillation}\label{sec:denoiz-distill}
We think that a model pre-trained with pseudo labels generated by clustering (e.g., k-means) outputs fine-grained clustered features; therefore, using the pre-trained model features as a training signal should be desirable.
HuBERT and WavLM create training signals for the next training iteration by clustering the features produced by the previous iteration model. However, since the previous model has already learned to target clustered labels, we think the model should produce clustered features.

That being said, we propose to combine denoising and distillation as a more flexible approach to learning from clustered targets. The distillation method typically forms a teacher-student network and uses the teacher's output as a training signal to train the student. Inspired by WavLM, we add noise to the student's input. In this case, the teacher's output acts as a microcluster center corresponding to each data sample, and the student learns to predict cluster centers regardless of the noise. As a result, the denoising distillation is expected to learn from fine-grained clustered features.

\subsection{Masked Modeling Duo for Speech}\label{sec:m2d-s}
Figure \ref{fig:system} shows M2D-S, which adds an offline network as a speech extension to M2D. We use two types of data (speech and background noise): speech only for the speech extension and noisy speech, a mixture of speech and noise, for M2D.

We distill the offline network as a teacher to M2D as a student.
In the speech extension, while the offline encoder produces features $h$, we concatenate $z_v$ and $\hat{z}_m$ from the output of M2D and project them using a linear layer to predict offline encoder features as $\hat{h}$.
In case M2D splits the input spectrogram in the frequency axis, the $\hat{h}$ is calculated to have one feature prediction per frame by concatenating the feature predictions for each frequency belonging to a time frame.

We calculate the offline loss $L_\text{off}$ by the MSE of \mbox{$l_2$-normalized} $h$ and $\hat{h}$, the same as Eq. \eqref{eq:eq-byol-mse} in M2D:
\vspace{-0.2cm}
\begin{equation}
L_\text{off} \triangleq ||l_2(h) - l_2(\hat{h})||^2_2 = 2 - 2 \cdot \frac{\langle h, \hat{h} \rangle }{||h||_2 \cdot ||\hat{h}||_2}
\label{eq:eq-off-loss}
\end{equation}
\vspace{-0.2cm}

The overall M2D-S loss $L_\text{m2dS}$ is then calculated by combining $L_\text{m2d}$ and $L_\text{off}$:
\vspace{-0.1cm}
\begin{equation}
L_\text{m2dS} = \lambda_\text{m2d} L_\text{m2d} + \lambda_\text{off} L_\text{off},
\label{eq:eq-m2ds-loss}
\end{equation}
where the loss weights $\lambda_\text{m2d}$ and $\lambda_\text{off}$ control the contribution.

The noisy speech is a mixture of background noise and speech in the $\alpha$ to $(1-\alpha)$ ratio, where $\alpha$ is a dataset noise ratio.

\section{Experiments}
First, we validate the effect of each modification made in M2D-S in ablation studies of the pre-training dataset (Section \ref{sec:exp-dataset}), the input size (Section \ref{sec:exp-input-shape}), and the pre-training tasks (Section \ref{sec:exp-abl-tasks}). Finally, we compare M2D-S combining all the best practices with SOTA models (Section \ref{sec:exp-sota}).

\subsection{Experimental Setup}\label{sec:exp-setup}
We used the same M2D configurations in M2D-S, including the use of ViT Base as the encoder, with a fixed masking ratio of 0.6.
For each experiment, the parameters of interest were varied with the following defaults:\footnote{\scriptsize{We provide complete details at: {\url{https://github.com/nttcslab/m2d/tree/master/speech}}}} dataset noise ratio $\alpha$ of 0.2, input duration $T$ of 2.08 s, and patch size of $80\times 4$.
M2D takes log-mel spectrogram features as input, which we preprocessed with a sampling frequency of 16,000 Hz, window size of 25 ms, hop size of 10 ms, and frequency bins $F$ of 80 in the range of 50 to 8,000 Hz and normalized using dataset statistics.

\vspace{0.1cm}
\noindent\textbf{Pre-training details}\hspace{0.2cm}
We set the number of epochs to 1,000 and the warm-up epochs to 60, where a single epoch consumes all LibriSpeech samples. All other settings were the same as in the M2D, including the batch size and optimizer settings.
We used the 9th transformer layer output of the 2nd-iteration HuBERT Base model\footnote{\scriptsize\url{https://huggingface.co/facebook/hubert-base-ls960}} as an offline encoder as in WavLM.

For the speech corpus, we used LibriSpeech with 281,241 samples (960 h) from all training splits (LS-960).
For the background noise dataset, we used AudioSet \cite{gemmeke2017audioset} (AS) with 2,005,132 samples (5,569 h) of 10-s audio from the balanced and unbalanced train segments.
We randomly sampled then randomly cropped a segment for the input duration and mixed it with the speech.
When mixing, the log-mel spectrograms were once reverted to a linear scale, mixed according to the ratio, and then converted back to log scale.

\vspace{0.1cm}
\noindent\textbf{Evaluation details}\hspace{0.2cm}
We evaluated all pre-trained models in SUPERB \cite{yang2021superb} for the speech task performance and a linear evaluation for the non-speech task. In the SUPERB evaluation, the model weights were frozen, and the weighted sum of the features from all transformer layers was used in the evaluation. The evaluation tasks include phoneme recognition (PR), keyword spotting (KS),  intent  classification (IC), speaker identification (SID), and emotion recognition (ER).

We used linear evaluation as a supplemental measure to assess non-speech task performance when specializing in speech.
The the linear evaluation and task details are the same as in BYOL-A \cite{niizumi2023byol-a}. We report the average accuracies of two environmental sound tasks (ENV) and three music tasks (MUS). In this evaluation, we trained a linear layer using only the final layer feature of the frozen model instead of using all layer features as in SUPERB.

\subsection{Pre-training Dataset Ablations}\label{sec:exp-dataset}
We assessed the impact of the dataset used for pre-training by comparing the performance when using the target domain dataset, i.e., LibriSpeech, and the general audio dataset, i.e., AudioSet.
To do so, we used M2D without speech extension and varied the dataset noise ratio $\alpha$.
When only AudioSet was used, we defined one epoch as the number of LibriSpeech samples, and we randomly sampled from AudioSet.

The results in Table \ref{tab:results-data-mix} show that the speech tasks perform best when only using LibriSpeech, and performance deteriorates as the ratio of AudioSet increases. Conversely, the non-speech tasks (ENV, MUS) perform best when only using AudioSet, and performance deteriorates as the ratio of LibriSpeech increases.
Exceptionally, SID performed best when the ratio $\alpha=0.2$ and there was background noise in the speech. However, the use of a speech domain dataset is generally considered to contribute to speech task performance.

\begin{table}[htb!]
\vspace{-5pt}
\caption{M2D pre-training dataset noise ratio ablations.\\
\footnotesize{(input duration $T=2.08s$ and patch size $80\times 4$)}}
\vspace{-5pt}
\label{tab:results-data-mix}
\centering
\resizebox{\columnwidth}{!}{%
\begin{tabular}{llllll|ll}
\toprule
 &    PR &     KS $^\star$ &     IC &     SID &    ER & \textcolor{black}{ENV}  & \textcolor{black}{MUS} \\
\vspace{-1pt} Dataset noise ratio $\alpha$       &  PER$\downarrow$ & Acc$\uparrow$& Acc$\uparrow$& Acc$\uparrow$& Acc$\uparrow$& \textcolor{black}{Acc$\uparrow$} & \textcolor{black}{Acc$\uparrow$} \\
\midrule
0.0 \textit{(LS-960 only)}  &\textbf{10.98}&  96.85 &\textbf{95.02}&  74.77 &\textbf{63.02}&\textcolor{black}{66.54}&\textcolor{black}{49.83}\\
0.1         &  11.97 &  97.06 &  94.38 &  77.89 &  61.44 &\textcolor{black}{72.77}&\textcolor{black}{51.68}\\
0.2         &  11.87 &  96.99 &  93.28 & \textbf{78.46} &  61.75 &\textcolor{black}{73.57}&\textcolor{black}{53.76}\\
0.3         &  12.19 &\textbf{97.23}&  94.46 &  78.25 &  61.55 &\textcolor{black}{74.67}&\textcolor{black}{54.44}\\
0.4         &  12.39 &  97.08 &  94.15 &  77.15 &  61.52 &\textcolor{black}{75.16}&\textcolor{black}{54.39}\\
0.5         &  12.53 &  96.82 &  92.14 &  76.58 &  61.07 &\textcolor{black}{76.38}&\textcolor{black}{54.73}\\
1.0 \textit{(AudioSet only)} &  27.04 &  95.60 &  82.78 &  68.54 &  60.85 &\textcolor{black}{\textbf{83.31}}&\textcolor{black}{\textbf{61.88}}\\
\bottomrule
\addlinespace[0.1cm]
\multicolumn{8}{l}{$^\star$ All KS results have been corrected. See Appendix \ref{sec:corrections} for the details.}\\
\end{tabular}
}
\vspace{-20pt}
\end{table}

\subsection{Input Size Ablations}\label{sec:exp-input-shape}
The patch splitting of the input spectrogram in the generic model is a key difference from the speech model.
In addition to patch size, we examined the impact on performance of different input durations using M2D without speech extension.

Table \ref{tab:exp-patch-size} patch size results indicate that the setting equivalent to the speech model ($80\times2$) has the best performance balance except for SID; for patch size $80\times2$ (Frequency $\times$ Time-step), 80 indicates no splitting on the frequency axis, and 2 corresponds to 20 ms per frame.
The results also show that the longer time steps (lower frame rates) degrade the performance, especially in PR, similar to the results in \cite{Meng2023Compressing}.

The patch sizes of $40\times4$ and $40\times2$, which assign two patches along the frequency axis, showed the best results for PR and SID, but inferior results on some other tasks. Overall, these results indicate that no patch split along the frequency axis provides a balanced better performance on the speech tasks.
Notably, our results align with SSAST \cite{gong2022ssast}.

The input duration results in Table \ref{tab:exp-input-dur} show a trend that the longer the input, the better the results. The trend of better performance when learning from features in longer series is consistent as it is in M2D.

\begin{table}[htb!]
\caption{M2D patch size ablations.\\
\footnotesize{(dataset noise ratio $\alpha=0.2$ and input duration $T=2.08s$)}}
\vspace{-5pt}
\label{tab:exp-patch-size}
\centering
\resizebox{\columnwidth}{!}{%
\begin{tabular}{llllll|ll}
\toprule
 &    PR &     KS $^\star$ &     IC &     SID &    ER & \textcolor{EMgray}{ENV}  & \textcolor{EMgray}{MUS} \\
\vspace{-1pt} Patch size $\textit{Freq.}\times \textit{Time}$      &  PER$\downarrow$ & Acc$\uparrow$& Acc$\uparrow$& Acc$\uparrow$& Acc$\uparrow$& \textcolor{EMgray}{Acc$\uparrow$} & \textcolor{EMgray}{Acc$\uparrow$} \\
\midrule
$16\times16$ (M2D \cite{niizumi2022M2D}) &  77.92 &  96.36 &  83.23 &  79.65 &  58.88 &\textcolor{EMgray}{80.27}&\textcolor{EMgray}{\textbf{59.60}}\\
$40\times8$  &  29.28 &  96.89 &  89.67 &  77.51 &  59.39 &\textcolor{EMgray}{78.56}&\textcolor{EMgray}{56.62}\\
$40\times4$  &\textbf{11.49}&  96.82 &  90.93 &  81.64 &  60.30 &\textcolor{EMgray}{80.00}&\textcolor{EMgray}{57.48}\\
$40\times2$  &  15.14 &  96.79 &  82.73 &\textbf{85.44}&  61.49 &\textcolor{EMgray}{\textbf{80.98}}&\textcolor{EMgray}{58.67}\\
$80\times8$  &  30.28 &  96.30 &  90.51 &  75.66 &  58.97 &\textcolor{EMgray}{72.62}&\textcolor{EMgray}{54.28}\\
$80\times4$  &  11.87 &  96.99 &  93.28 &  78.46 &\textbf{61.75}&\textcolor{EMgray}{73.57}&\textcolor{EMgray}{53.76}\\
$80\times2$ ($\Leftrightarrow$ speech models) &  11.74 &\textbf{97.25}&\textbf{93.51}&  78.86 &  60.67 &\textcolor{EMgray}{74.98}&\textcolor{EMgray}{53.28}\\
\bottomrule
\addlinespace[0.1cm]
\multicolumn{8}{l}{$^\star$ All KS results have been corrected. See Appendix \ref{sec:corrections} for the details.}\\
\end{tabular}
}
\vspace{-10pt}
\end{table}

\begin{table}[htb!]
\vspace{-5pt}
\caption{M2D input duration ablations.\\
\footnotesize{(dataset noise ratio $\alpha=0.2$ and patch size $80\times 4$)}}
\vspace{-5pt}
\label{tab:exp-input-dur}
\centering
\resizebox{\columnwidth}{!}{%
\begin{tabular}{llllll|ll}
\toprule
 &    PR &     KS $^\star$ &     IC &     SID &    ER & \textcolor{EMgray}{ENV}  & \textcolor{EMgray}{MUS} \\
\vspace{-1pt} Input duration $T$   &  PER$\downarrow$ & Acc$\uparrow$& Acc$\uparrow$& Acc$\uparrow$& Acc$\uparrow$& \textcolor{EMgray}{Acc$\uparrow$} & \textcolor{EMgray}{Acc$\uparrow$} \\
\midrule
T=2.08s &  11.87 &  96.99 &  93.28 &  78.46 &  61.75 &\textcolor{EMgray}{73.57}&\textcolor{EMgray}{\textbf{53.76}}\\
T=3.04s &   9.81 &  97.09 &  95.15 &  79.18 &  63.39 &\textcolor{EMgray}{74.36}&\textcolor{EMgray}{52.14}\\
T=4.00s &   8.50 &\textbf{97.34}&  94.83 &\textbf{81.24}&  63.81 &\textcolor{EMgray}{74.11}&\textcolor{EMgray}{52.23}\\
T=5.12s &   8.10 &  97.17 &  94.70 &  78.73 &\textbf{65.47}&\textcolor{EMgray}{\textbf{74.69}}&\textcolor{EMgray}{51.66}\\
T=6.08s &\textbf{7.74}&  97.17 &\textbf{95.50}&  80.48 &  64.06 &\textcolor{EMgray}{72.42}&\textcolor{EMgray}{50.64}\\
\bottomrule
\addlinespace[0.1cm]
\multicolumn{8}{l}{$^\star$ All KS results have been corrected. See Appendix \ref{sec:corrections} for the details.}\\
\end{tabular}
}
\vspace{-10pt}
\end{table}

\begin{table}[htb!]
\vspace{-5pt}
\caption{M2D-S pre-training task ablations.\\
\footnotesize{(input duration $T=2.08s$, and patch size $80\times 4$)}}
\vspace{-5pt}
\label{tab:exp-tasks}
\centering
\resizebox{\columnwidth}{!}{%
\begin{tabular}{lllllllll|ll}
\toprule
&\multicolumn{2}{c}{Offline} & Online &    PR &     KS $^\star$ &     IC &     SID &    ER & \textcolor{EMgray}{ENV}  & \textcolor{EMgray}{MUS} \\
 \cmidrule(lr){2-3} \cmidrule(lr){4-4}
&\vspace{-1pt} denoise & distill. & M2D &  PER$\downarrow$ & Acc$\uparrow$& Acc$\uparrow$& Acc$\uparrow$& Acc$\uparrow$& \textcolor{EMgray}{Acc$\uparrow$} & \textcolor{EMgray}{Acc$\uparrow$} \\
\midrule
(a)&&&\checkmark        &  11.87 &\textbf{96.99}&  93.28 &\textbf{78.46}&  61.75 &\textcolor{EMgray}{\textbf{73.57}}&\textcolor{EMgray}{\textbf{53.76}}\\
(b)&&\checkmark&           &   8.10 &  96.05 &  94.81 &  65.51 &  61.27 &\textcolor{EMgray}{56.96}&\textcolor{EMgray}{44.16}\\
(c)&\checkmark&\checkmark& &   7.80 &  94.66 &  93.38 &  70.03 &  61.58 &\textcolor{EMgray}{42.42}&\textcolor{EMgray}{35.36}\\
(d)&&\checkmark&\checkmark                      &   8.78 &  95.72 &  91.35 &  62.68 &  60.85 &\textcolor{EMgray}{57.15}&\textcolor{EMgray}{43.75}\\
(e)&\checkmark&\checkmark&\checkmark      &\textbf{7.02}&  95.48 &\textbf{97.10}&  78.12 &\textbf{64.09}&\textcolor{EMgray}{55.49}&\textcolor{EMgray}{41.51}\\
\bottomrule
\addlinespace[0.1cm]
\multicolumn{11}{l}{$^\star$ All KS results have been corrected. See Appendix \ref{sec:corrections} for the details.}\\
\end{tabular}
}
\vspace{-10pt}
\end{table}

\subsection{Pre-training Task Ablations}\label{sec:exp-abl-tasks}
We assessed the effectiveness of the denoising distillation and the M2D task and also tested without denoising.
We switched the tasks in the offline and M2D networks by setting $\lambda_\text{off}$ and $\lambda_\text{m2d}$ in Eq. \eqref{eq:eq-m2ds-loss} to 0 or 1.0, respectively, and switched the denoising task by setting $\alpha$ to 0 (disabled) or 0.2 (enabled).

The results in Table \ref{tab:exp-tasks} show that (b) distillation only is better than (a) M2D or (d) M2D plus distillation, indicating that the distillation of a speech model trained with clustered features is effective for the speech tasks.
Compared to (b) with only distillation, (c) with additional denoising performs better except for KS and IC, indicating the effectiveness of the denoising task.
M2D-S configuration (e), learning all tasks together, significantly improves task performance, showing that combining denoising distillation and M2D is remarkably effective.


\begin{table}[tb!]
\caption{Comparison with SOTA speech models. \\
\footnotesize{($\lambda_\text{off}=0.5$, $\lambda_\text{m2d}=1$, $\alpha=0.2$, and patch size $80\times 2$)}}
\vspace{-5pt}
\label{tab:results-sota}
\centering
\resizebox{\columnwidth}{!}{%
\begin{tabular}{lllllll|ll}
\toprule
 & &    PR &     KS $^\star$ &     IC &     SID &    ER & \textcolor{EMgray}{ENV}  & \textcolor{EMgray}{MUS} \\
\vspace{-1pt} Model     & Dataset &  PER$\downarrow$ & Acc$\uparrow$& Acc$\uparrow$& Acc$\uparrow$& Acc$\uparrow$& \textcolor{EMgray}{Acc$\uparrow$} & \textcolor{EMgray}{Acc$\uparrow$} \\
\midrule
wav2vec2.0 Base \cite{baevski2020wav2vec2}$^{\dagger}$ & \footnotesize{LS-960} & 5.74 & 96.23 & 92.35 & 75.18 & 63.43 & \textcolor{EMgray}{\textit{37.66}} &  \textcolor{EMgray}{\textit{32.02}} \\
HuBERT Base \cite{Hsu2021HuBERT}$^{\dagger}$ & \footnotesize{LS-960} & 5.41 & 96.30 & 98.34 & 81.42 & 64.92 & \textcolor{EMgray}{\textit{62.76}} & \textcolor{EMgray}{\textit{46.26}} \\
WavLM Base \cite{Chen2022WavLM}$^{\dagger}$ & \scriptsize{LS-960+DNS} & \textbf{4.84} & 96.79 & \textbf{98.63} & \textbf{84.51} & 65.94 & \textcolor{EMgray}{\textit{54.45}} & \textcolor{EMgray}{\textit{40.98}} \\
\midrule
\multicolumn{6}{l}{\textit{(Proposed using ViT \textit{Base})}} &&& \\
M2D-S T=4.0s  & \footnotesize{LS-960+AS} &   5.72 &  96.47 &  97.80 &  81.97 &  \textbf{66.36} &\textcolor{EMgray}{53.22}&\textcolor{EMgray}{41.71}\\
M2D-S T=5.12s   & \footnotesize{LS-960+AS} &   5.64 &  \textbf{96.87} &  97.65 &  80.69 &  65.35 &\textcolor{EMgray}{57.34}&\textcolor{EMgray}{43.23}\\

M2D-S T=6.08s & \footnotesize{LS-960+AS} &   5.33 & 96.80 &  97.63 &  81.74 &    66.13 &\textcolor{EMgray}{54.77}&\textcolor{EMgray}{43.75}\\

\multicolumn{6}{l}{\textit{(Conventional using ViT \textit{Base})}} &&& \\
M2D ratio=0.6 \cite{niizumi2022M2D}$^{\natural}$ & \scriptsize{AS} &  78.30 &  95.65 &  76.77 &  80.68 &  61.17 &  \textcolor{EMgray}{\textbf{88.63}}&\textcolor{EMgray}{\textbf{66.56}}\\
\midrule
\multicolumn{6}{l}{\textit{(Reference \textit{Large} models)}} &&& \\
\textcolor{gray}{wav2vec\small{2.0 Large \cite{baevski2020wav2vec2}}$^{\dagger}$} & \textcolor{gray}{\footnotesize{LL-60k}} & \textcolor{gray}{4.75} & \textcolor{gray}{96.66} & \textcolor{gray}{95.28} & \textcolor{gray}{86.14} & \textcolor{gray}{65.64} & \textcolor{gray}{\textit{60.82}} &  \textcolor{gray}{\textit{42.75}} \\
\textcolor{gray}{HuBERT Large \cite{Hsu2021HuBERT}$^{\dagger}$} & \textcolor{gray}{\footnotesize{LL-60k}} & \textcolor{gray}{3.53} & \textcolor{gray}{95.29} & \textcolor{gray}{98.76} & \textcolor{gray}{90.33} & \textcolor{gray}{67.62} & \textcolor{gray}{\textit{59.51}} &  \textcolor{gray}{\textit{44.35}} \\
\textcolor{gray}{WavLM Large \cite{Chen2022WavLM}$^{\dagger}$} & \textcolor{gray}{\footnotesize{Mix-94k}} & \textcolor{gray}{\textbf{3.06}} & \textcolor{gray}{97.86} & \textcolor{gray}{\textbf{99.31}} & \textcolor{gray}{\textbf{95.49}} & \textcolor{gray}{\textbf{70.62}} & \textcolor{gray}{\textit{69.32}} &  \textcolor{gray}{\textit{50.56}} \\
\bottomrule
\addlinespace[0.1cm]
\multicolumn{9}{l}{$^{\dagger}$
ENV and MUS results were obtained using publicly available pre-trained models.}\\
\multicolumn{9}{l}{$^{\natural}$
The original M2D takes input with $T=6.08s$ and uses a patch size of $16\times 16$.}\\
\multicolumn{9}{l}{$^\star$ All KS results have been corrected. See Appendix \ref{sec:corrections} for the details.}\\
\end{tabular}
}
\vspace{-15pt}
\end{table}

\subsection{Comparison with SOTA}\label{sec:exp-sota}
While previous experiments have improved performance, they have not reached SOTA performance; therefore, we combine all the settings that offered the best performance in M2D-S.
Table \ref{tab:results-sota} compares the original M2D, proposed M2D-S, and the SOTA speech \textit{Base} models, which have a number of parameters close to that in M2D. We tested three input durations $T$ of M2D-S while keeping the other settings fixed. We also optimized $\lambda_\text{off}$ and $\lambda_\text{m2d}$ by parameter search.

The results show that M2D-S significantly improves the performance of M2D to a level comparable to SOTA speech models. Furthermore, M2D-S outperforms these SOTA models on KS and ER tasks. Notably, this is achieved without a CNN acoustic feature extractor used by the previous methods, while M2D-S takes log-mel spectrograms as input.
On the other hand, M2D-S performed worse on the non-speech ENV and MUS tasks than M2D and HuBERT, which is interestingly similar to WavLM. Both WavLM and M2D-S learn speech denoising, which we suspect may be the cause of this side effect.

\section{Discussion for Future Specialization}
This study explored the requirements for specializing M2D in speech, and the lessons learned may be useful for specialization in other acoustic fields.
\begin{itemize}
\item Optimizing the patch size may be effective for other applications. While splitting only the time axis was effective for speech, splitting along both the frequency and time axes has been reported as effective for other audio tasks \cite{gong2022ssast, niizumi2022msm-mae}.
\item In addition to using a dataset from the target domain, combining a denoising task with background noise may be effective in learning application-focused representations.
\item Learning the denoising distillation task requires an appropriate teacher model. Although we used a model that learned pseudo-labels by clustering, pre-trained classifier models available in the application field may be effective since they learn from class labels, considered human-curated clusters.
\end{itemize}



\section{Conclusion}
This paper explored the challenges and solutions in specializing a general-purpose audio representation (generic model) to a specific domain using speech as an example.
We found the challenge is to incorporate speech-specific techniques, such as learning from clustered features.

To achieve speech SOTA performance, we proposed a new task, denoising distillation, to learn from fine-grained clustered features, and M2D for Speech (M2D-S), which jointly learns the denoising distillation task and M2D masked prediction task. In the experiments on the SUPERB benchmark, M2D-S significantly improved the performance of M2D. It performed comparable to SOTA speech models (e.g., HuBERT and WavLM) and outperformed them in keyword spotting and emotion recognition tasks, demonstrating that a generic model can specialize and be useful in a challenging field.

Our study indicated that a generic model is potentially useful for specializing in other audio applications. Our code is available online for future studies.

\bibliographystyle{IEEEtran}
\bibliography{refs}

\begin{thebibliography}{10}
\providecommand{\url}[1]{#1}
\csname url@samestyle\endcsname
\providecommand{\newblock}{\relax}
\providecommand{\bibinfo}[2]{#2}
\providecommand{\BIBentrySTDinterwordspacing}{\spaceskip=0pt\relax}
\providecommand{\BIBentryALTinterwordstretchfactor}{4}
\providecommand{\BIBentryALTinterwordspacing}{\spaceskip=\fontdimen2\font plus
\BIBentryALTinterwordstretchfactor\fontdimen3\font minus
  \fontdimen4\font\relax}
\providecommand{\BIBforeignlanguage}[2]{{%
\expandafter\ifx\csname l@#1\endcsname\relax
\typeout{** WARNING: IEEEtran.bst: No hyphenation pattern has been}%
\typeout{** loaded for the language `#1'. Using the pattern for}%
\typeout{** the default language instead.}%
\else
\language=\csname l@#1\endcsname
\fi
#2}}
\providecommand{\BIBdecl}{\relax}
\BIBdecl

\bibitem{gong2022ssast}
Y.~Gong, C.-I. Lai, Y.-A. Chung, and J.~Glass, ``{SSAST: Self-Supervised Audio
  Spectrogram Transformer},'' in \emph{AAAI}, vol.~36, no.~10, 2022, pp.
  10\,699--10\,709.

\bibitem{Baade2022MAE-AST}
A.~Baade, P.~Peng, and D.~Harwath, ``{MAE-AST: Masked Autoencoding Audio
  Spectrogram Transformer},'' in \emph{Interspeech}, 2022, pp. 2438--2442.

\bibitem{niizumi2022msm-mae}
D.~Niizumi, D.~Takeuchi, Y.~Ohishi, N.~Harada, and K.~Kashino, ``{Masked
  Spectrogram Modeling using Masked Autoencoders for Learning General-purpose
  Audio Representation},'' in \emph{HEAR: Holistic Evaluation of Audio
  Representations (NeurIPS 2021 Competition)}, vol. 166, 2022, pp. 1--24.

\bibitem{niizumi2022M2D}
------, ``{Masked Modeling Duo: Learning Representations by Encouraging Both
  Networks to Model the Input},'' in \emph{ICASSP}, 2023, pp. 1--5.

\bibitem{chen2022beats}
S.~Chen, Y.~Wu, C.~Wang, S.~Liu, D.~Tompkins, Z.~Chen, and F.~Wei, ``{BEATs:
  Audio Pre-Training with Acoustic Tokenizers},'' in \emph{ICML}, 2023.

\bibitem{LIU2022ASurvey}
S.~Liu, A.~Mallol-Ragolta, E.~Parada-Cabaleiro, K.~Qian, X.~Jing, A.~Kathan,
  B.~Hu, and B.~W. Schuller, ``Audio self-supervised learning: A survey,''
  \emph{Patterns}, vol.~3, no.~12, p. 100616, 2022.

\bibitem{baevski2020wav2vec2}
A.~Baevski, Y.~Zhou, A.~Mohamed, and M.~Auli, ``wav2vec 2.0: {A} framework for
  self-supervised learning of speech representations,'' in \emph{NeurIPS},
  2020.

\bibitem{Hsu2021HuBERT}
W.-N. Hsu, B.~Bolte, Y.-H.~H. Tsai, K.~Lakhotia, R.~Salakhutdinov, and
  A.~Mohamed, ``{HuBERT: Self-Supervised Speech Representation Learning by
  Masked Prediction of Hidden Units},'' \emph{IEEE/ACM Trans. Audio, Speech,
  Language Process.}, p. 3451–3460, 2021.

\bibitem{Chen2022WavLM}
S.~Chen, C.~Wang, Z.~Chen, Y.~Wu, S.~Liu, Z.~Chen, J.~Li, N.~Kanda,
  T.~Yoshioka, X.~Xiao, J.~Wu, L.~Zhou, S.~Ren, Y.~Qian, Y.~Qian, J.~Wu,
  M.~Zeng, X.~Yu, and F.~Wei, ``{WavLM: Large-Scale Self-Supervised
  Pre-Training for Full Stack Speech Processing},'' \emph{IEEE J. Sel. Top.
  Signal Process.}, vol.~16, no.~6, p. 1505–1518, 2022.

\bibitem{yang2021superb}
S.~wen Yang, P.-H. Chi, Y.-S. Chuang, C.-I.~J. Lai, K.~Lakhotia, Y.~Y. Lin,
  A.~T. Liu, J.~Shi, X.~Chang, G.-T. Lin, T.-H. Huang, W.-C. Tseng, K.~tik Lee,
  D.-R. Liu, Z.~Huang, S.~Dong, S.-W. Li, S.~Watanabe, A.~Mohamed, and
  H.~yi~Lee, ``{SUPERB: Speech Processing Universal PERformance Benchmark},''
  in \emph{Interspeech}, 2021, pp. 1194--1198.

\bibitem{Liu2020Mockingjay}
A.~T. Liu, S.-w. Yang, P.-H. Chi, P.-c. Hsu, and H.-y. Lee, ``{Mockingjay:
  Unsupervised Speech Representation Learning with Deep Bidirectional
  Transformer Encoders},'' in \emph{ICASSP}, 2020, pp. 6419--6423.

\bibitem{Liu2021TERA}
A.~T. Liu, S.-W. Li, and H.-y. Lee, ``{TERA: Self-Supervised Learning of
  Transformer Encoder Representation for Speech},'' \emph{IEEE/ACM Trans.
  Audio, Speech, Language Process.}, vol.~29, pp. 2351--2366, 2021.

\bibitem{zhang2022bigssl}
Y.~Zhang, D.~S. Park, W.~Han, J.~Qin, A.~Gulati, J.~Shor, A.~Jansen, Y.~Xu,
  Y.~Huang, S.~Wang, Z.~Zhou, B.~Li, M.~Ma, W.~Chan, J.~Yu, Y.~Wang, L.~Cao,
  K.~C. Sim, B.~Ramabhadran, T.~N. Sainath, F.~Beaufays, Z.~Chen, Q.~V. Le,
  C.-C. Chiu, R.~Pang, and Y.~Wu, ``{BigSSL}: Exploring the frontier of
  large-scale semi-supervised learning for automatic speech recognition,''
  \emph{IEEE J. Sel. Top. Signal Process.}, vol.~16, no.~6, p. 1519–1532,
  2022.

\bibitem{baevski2022data2vec}
A.~Baevski, W.-N. Hsu, Q.~Xu, A.~Babu, J.~Gu, and M.~Auli, ``data2vec: A
  general framework for self-supervised learning in speech, vision and
  language,'' in \emph{ICML}, 2022, pp. 1298--1312.

\bibitem{gong2021ast}
Y.~Gong, Y.-A. Chung, and J.~Glass, ``{AST: Audio Spectrogram Transformer},''
  in \emph{Interspeech}, 2021, pp. 571--575.

\bibitem{Li2022ATST}
X.~LI and X.~Li, ``{ATST: Audio Representation Learning with Teacher-Student
  Transformer},'' in \emph{Interspeech}, 2022, pp. 4172--4176.

\bibitem{huang2022amae}
P.-Y. Huang, H.~Xu, J.~Li, A.~Baevski, M.~Auli, W.~Galuba, F.~Metze, and
  C.~Feichtenhofer, ``Masked autoencoders that listen,'' in \emph{NeurIPS},
  2022.

\bibitem{ViT}
A.~Dosovitskiy, L.~Beyer, A.~Kolesnikov, D.~Weissenborn, X.~Zhai,
  T.~Unterthiner, M.~Dehghani, M.~Minderer, G.~Heigold, S.~Gelly, J.~Uszkoreit,
  and N.~Houlsby, ``An image is worth 16x16 words: Transformers for image
  recognition at scale,'' in \emph{ICLR}, 2021.

\bibitem{le2023lungsound}
L.~Melms, R.~R. Ilesan, U.~Köhler, O.~Hildebrandt, R.~Conradt, J.~Eckstein,
  C.~Atila, S.~Matrood, B.~Schieffer, J.~R. Schaefer, T.~Müller,
  J.~Obergassel, N.~Schlicker, and M.~C. Hirsch, ``Training one model to detect
  heart and lung sound events from single point auscultations,'' \emph{arXiv
  preprint arXiv:2301.06078}, 2023.

\bibitem{elbanna22BYOL-S}
G.~Elbanna, N.~Scheidwasser-Clow, M.~Kegler, P.~Beckmann, K.~El~Hajal, and
  M.~Cernak, ``{BYOL-S}: Learning self-supervised speech representations by
  bootstrapping,'' in \emph{HEAR: Holistic Evaluation of Audio Representations
  (NeurIPS 2021 Competition)}, vol. 166, 2022, pp. 25--47.

\bibitem{niizumi2023byol-a}
D.~Niizumi, D.~Takeuchi, Y.~Ohishi, N.~Harada, and K.~Kashino, ``{BYOL for
  Audio}: Exploring pre-trained general-purpose audio representations,''
  \emph{IEEE/ACM Trans. Audio, Speech, Language Process.}, vol.~31, p.
  137–151, 2023.

\bibitem{lauscher-etal-2020-specializing}
A.~Lauscher, I.~Vuli{\'c}, E.~M. Ponti, A.~Korhonen, and G.~Glava{\v{s}},
  ``Specializing unsupervised pretraining models for word-level semantic
  similarity,'' in \emph{COLING}, 2020, pp. 1371--1383.

\bibitem{bert}
J.~Devlin, M.~Chang, K.~Lee, and K.~Toutanova, ``{BERT:} pre-training of deep
  bidirectional transformers for language understanding,'' in \emph{NAACL-HLT},
  2019, pp. 4171--4186.

\bibitem{DistilHuBERT}
H.-J. Chang, S.-w. Yang, and H.-y. Lee, ``{DistilHuBERT}: Speech representation
  learning by layer-wise distillation of hidden-unit bert,'' in \emph{ICASSP},
  2022, pp. 7087--7091.

\bibitem{ma2022mt4ssl}
Z.~Ma, Z.~Zhen, C.~Tang, Y.~Wang, and X.~Chen, ``{MT4SSL: Boosting
  Self-Supervised Speech Representation Learning by Integrating Multiple
  Targets},'' \emph{to appear at Interspeech}, 2023.

\bibitem{guimares2023robustdistiller}
H.~R. Guimarães, A.~Pimentel, A.~R. Avila, M.~Rezagholizadeh, B.~Chen, and
  T.~H. Falk, ``{RobustDistiller: Compressing Universal Speech Representations
  for Enhanced Environment Robustness},'' in \emph{ICASSP}, 2023, pp. 1--5.

\bibitem{Panayotov2015LibrispeechAA}
V.~Panayotov, G.~Chen, D.~Povey, and S.~Khudanpur, ``Librispeech: An asr corpus
  based on public domain audio books,'' in \emph{ICASSP}, 2015, pp. 5206--5210.

\bibitem{gemmeke2017audioset}
J.~F. Gemmeke, D.~P.~W. Ellis, D.~Freedman, A.~Jansen, W.~Lawrence, R.~C.
  Moore, M.~Plakal, and M.~Ritter, ``{Audio Set}: An ontology and human-labeled
  dataset for audio events,'' in \emph{ICASSP}, 2017, pp. 776--780.

\bibitem{Meng2023Compressing}
Y.~Meng, H.-J. Chen, J.~Shi, S.~Watanabe, P.~Garcia, H.-y. Lee, and H.~Tang,
  ``On compressing sequences for self-supervised speech models,'' in
  \emph{SLT}, 2022, pp. 1128--1135.

\bibitem{radford2022whisper}
A.~Radford, J.~W. Kim, T.~Xu, G.~Brockman, C.~McLeavey, and I.~Sutskever,
  ``{Robust Speech Recognition via Large-Scale Weak Supervision},'' \emph{arXiv
  preprint arXiv:2212.04356}, 2022.

\bibitem{chemudupati2023transferwhisper}
V.~Chemudupati, M.~Tahaei, H.~Guimaraes, A.~Pimentel, A.~Avila,
  M.~Rezagholizadeh, B.~Chen, and T.~Falk, ``On the transferability of
  whisper-based representations for "in-the-wild" cross-task downstream speech
  applications,'' \emph{to appear at Interspeech}, 2023.

\end{thebibliography}

\newpage

\appendix


\section{Corrections} \label{sec:corrections}

In the Interspeech2023 paper, we identified errors in the aggregation of KS task results. Consequently, all KS task results in Tables 2 through 6 have been corrected in this paper.

Tables \ref{corr:tab:results-data-mix} through \ref{corr:tab:results-sota} show the difference side by side.
We apologize for the errors and any confusion they may have caused.

\begin{table}[htbp!]
\caption{Corrections of Table 2: M2D pre-training dataset noise ratio ablations.\\
\footnotesize{(input duration $T=2.08s$ and patch size $80\times 4$)}}
\label{corr:tab:results-data-mix}
\vspace{-5pt}
\begin{subtable}[t]{0.49\textwidth}
\centering
\resizebox{\textwidth}{!}{%
\begin{tabular}{llllll|ll}
\toprule
 &    PR &     KS &     IC &     SID &    ER & \textcolor{black}{ENV}  & \textcolor{black}{MUS} \\
\vspace{-1pt} Dataset noise ratio $\alpha$       &  PER$\downarrow$ & Acc$\uparrow$& Acc$\uparrow$& Acc$\uparrow$& Acc$\uparrow$& \textcolor{black}{Acc$\uparrow$} & \textcolor{black}{Acc$\uparrow$} \\
\midrule
0.0 \textit{(LS-960 only)}  &\textbf{10.98}&\textcolor{red}{\textbf{96.20}}&\textbf{95.02}&  74.77 &\textbf{63.02}&\textcolor{black}{66.54}&\textcolor{black}{49.83}\\
0.1         &  11.97 &  \textcolor{red}{91.54} &  94.38 &  77.89 &  61.44 &\textcolor{black}{72.77}&\textcolor{black}{51.68}\\
0.2         &  11.87 &  \textcolor{red}{93.30} &  93.28 & \textbf{78.46} &  61.75 &\textcolor{black}{73.57}&\textcolor{black}{53.76}\\
0.3         &  12.19 &  \textcolor{red}{93.38} &  94.46 &  78.25 &  61.55 &\textcolor{black}{74.67}&\textcolor{black}{54.44}\\
0.4         &  12.39 &  \textcolor{red}{95.15} &  94.15 &  77.15 &  61.52 &\textcolor{black}{75.16}&\textcolor{black}{54.39}\\
0.5         &  12.53 &  \textcolor{red}{93.70} &  92.14 &  76.58 &  61.07 &\textcolor{black}{76.38}&\textcolor{black}{54.73}\\
1.0 \textit{(AudioSet only)} &  27.04 &  \textcolor{red}{81.91} &  82.78 &  68.54 &  60.85 &\textcolor{black}{\textbf{83.31}}&\textcolor{black}{\textbf{61.88}}\\
\bottomrule\\
\end{tabular}
}
\vspace{-8pt}
\caption{Original table}
\label{original:tab:results-data-mix}
\end{subtable}
\hfill
\begin{subtable}[t]{0.49\textwidth}
\centering
\resizebox{\textwidth}{!}{%
\begin{tabular}{llllll|ll}
\toprule
 &    PR &     KS &     IC &     SID &    ER & \textcolor{black}{ENV}  & \textcolor{black}{MUS} \\
\vspace{-1pt} Dataset noise ratio $\alpha$       &  PER$\downarrow$ & Acc$\uparrow$& Acc$\uparrow$& Acc$\uparrow$& Acc$\uparrow$& \textcolor{black}{Acc$\uparrow$} & \textcolor{black}{Acc$\uparrow$} \\
\midrule
0.0 \textit{(LS-960 only)}  &\textbf{10.98}&  \textcolor{red}{96.85} &\textbf{95.02}&  74.77 &\textbf{63.02}&\textcolor{black}{66.54}&\textcolor{black}{49.83}\\
0.1         &  11.97 &  \textcolor{red}{97.06} &  94.38 &  77.89 &  61.44 &\textcolor{black}{72.77}&\textcolor{black}{51.68}\\
0.2         &  11.87 &  \textcolor{red}{96.99} &  93.28 & \textbf{78.46} &  61.75 &\textcolor{black}{73.57}&\textcolor{black}{53.76}\\
0.3         &  12.19 &\textcolor{red}{\textbf{97.23}}&  94.46 &  78.25 &  61.55 &\textcolor{black}{74.67}&\textcolor{black}{54.44}\\
0.4         &  12.39 &  \textcolor{red}{97.08} &  94.15 &  77.15 &  61.52 &\textcolor{black}{75.16}&\textcolor{black}{54.39}\\
0.5         &  12.53 &  \textcolor{red}{96.82} &  92.14 &  76.58 &  61.07 &\textcolor{black}{76.38}&\textcolor{black}{54.73}\\
1.0 \textit{(AudioSet only)} &  27.04 &  \textcolor{red}{95.60} &  82.78 &  68.54 &  60.85 &\textcolor{black}{\textbf{83.31}}&\textcolor{black}{\textbf{61.88}}\\
\bottomrule\\
\end{tabular}
}
\vspace{-8pt}
\caption{Corrected table}
\label{ccorrected:tab:results-data-mix}
\hfill
\end{subtable}
\vspace{-10pt}
\end{table}

\begin{table}[htbp!]
\vspace{-10pt}
\caption{Corrections of Table 3: M2D patch size ablations.\\
\footnotesize{(dataset noise ratio $\alpha=0.2$ and input duration $T=2.08s$)}}
\label{corr:tab:exp-patch-size}
\vspace{-5pt}
\begin{subtable}[t]{0.49\textwidth}
\centering
\resizebox{\columnwidth}{!}{%
\begin{tabular}{llllll|ll}
\toprule
 &    PR &     KS &     IC &     SID &    ER & \textcolor{EMgray}{ENV}  & \textcolor{EMgray}{MUS} \\
\vspace{-1pt} Patch size $\textit{Freq.}\times \textit{Time}$      &  PER$\downarrow$ & Acc$\uparrow$& Acc$\uparrow$& Acc$\uparrow$& Acc$\uparrow$& \textcolor{EMgray}{Acc$\uparrow$} & \textcolor{EMgray}{Acc$\uparrow$} \\
\midrule
$16\times16$ (M2D \cite{niizumi2022M2D}) &  77.92 &  \textcolor{red}{86.24} &  83.23 &  79.65 &  58.88 &\textcolor{EMgray}{80.27}&\textcolor{EMgray}{\textbf{59.60}}\\
$40\times8$  &  29.28 &  \textcolor{red}{90.25} &  89.67 &  77.51 &  59.39 &\textcolor{EMgray}{78.56}&\textcolor{EMgray}{56.62}\\
$40\times4$  &\textbf{11.49}&  \textcolor{red}{91.77} &  90.93 &  81.64 &  60.30 &\textcolor{EMgray}{80.00}&\textcolor{EMgray}{57.48}\\
$40\times2$  &  15.14 &  \textcolor{red}{81.47} &  82.73 &\textbf{85.44}&  61.49 &\textcolor{EMgray}{\textbf{80.98}}&\textcolor{EMgray}{58.67}\\
$80\times8$  &  30.28 &  \textcolor{red}{89.69} &  90.51 &  75.66 &  58.97 &\textcolor{EMgray}{72.62}&\textcolor{EMgray}{54.28}\\
$80\times4$  &  11.87 &\textcolor{red}{\textbf{93.30}}&  93.28 &  78.46 &\textbf{61.75}&\textcolor{EMgray}{73.57}&\textcolor{EMgray}{53.76}\\
$80\times2$ ($\Leftrightarrow$ speech models) &  11.74 &  \textcolor{red}{93.28} &\textbf{93.51}&  78.86 &  60.67 &\textcolor{EMgray}{74.98}&\textcolor{EMgray}{53.28}\\
\bottomrule\\
\end{tabular}
}
\vspace{-8pt}
\caption{Original table}
\label{original:tab:exp-patch-size}
\end{subtable}
\hfill
\begin{subtable}[t]{0.49\textwidth}
\centering
\resizebox{\textwidth}{!}{%
\begin{tabular}{llllll|ll}
\toprule
 &    PR &     KS &     IC &     SID &    ER & \textcolor{EMgray}{ENV}  & \textcolor{EMgray}{MUS} \\
\vspace{-1pt} Patch size $\textit{Freq.}\times \textit{Time}$      &  PER$\downarrow$ & Acc$\uparrow$& Acc$\uparrow$& Acc$\uparrow$& Acc$\uparrow$& \textcolor{EMgray}{Acc$\uparrow$} & \textcolor{EMgray}{Acc$\uparrow$} \\
\midrule
$16\times16$ (M2D \cite{niizumi2022M2D}) &  77.92 &  \textcolor{red}{96.36} &  83.23 &  79.65 &  58.88 &\textcolor{EMgray}{80.27}&\textcolor{EMgray}{\textbf{59.60}}\\
$40\times8$  &  29.28 &  \textcolor{red}{96.89} &  89.67 &  77.51 &  59.39 &\textcolor{EMgray}{78.56}&\textcolor{EMgray}{56.62}\\
$40\times4$  &\textbf{11.49}&  \textcolor{red}{96.82} &  90.93 &  81.64 &  60.30 &\textcolor{EMgray}{80.00}&\textcolor{EMgray}{57.48}\\
$40\times2$  &  15.14 &  \textcolor{red}{96.79} &  82.73 &\textbf{85.44}&  61.49 &\textcolor{EMgray}{\textbf{80.98}}&\textcolor{EMgray}{58.67}\\
$80\times8$  &  30.28 &  \textcolor{red}{96.30} &  90.51 &  75.66 &  58.97 &\textcolor{EMgray}{72.62}&\textcolor{EMgray}{54.28}\\
$80\times4$  &  11.87 &  \textcolor{red}{96.99} &  93.28 &  78.46 &\textbf{61.75}&\textcolor{EMgray}{73.57}&\textcolor{EMgray}{53.76}\\
$80\times2$ ($\Leftrightarrow$ speech models) &  11.74 &\textcolor{red}{\textbf{97.25}}&\textbf{93.51}&  78.86 &  60.67 &\textcolor{EMgray}{74.98}&\textcolor{EMgray}{53.28}\\
\bottomrule\\
\end{tabular}
}
\vspace{-8pt}
\caption{Corrected table}
\label{ccorrected:tab:exp-patch-size}
\hfill
\end{subtable}
\end{table}

\begin{table}[tbp!]
\vspace{-20pt}
\caption{Corrections of Table 4: M2D input duration ablations.
\footnotesize{(dataset noise ratio $\alpha=0.2$ and patch size $80\times 4$)}}
\label{corr:tab:exp-input-dur}
\vspace{-5pt}
\begin{subtable}[t]{0.49\textwidth}
\centering
\resizebox{\columnwidth}{!}{%
\begin{tabular}{llllll|ll}
\toprule
 &    PR &     KS &     IC &     SID &    ER & \textcolor{EMgray}{ENV}  & \textcolor{EMgray}{MUS} \\
\vspace{-1pt} Input duration $T$   &  PER$\downarrow$ & Acc$\uparrow$& Acc$\uparrow$& Acc$\uparrow$& Acc$\uparrow$& \textcolor{EMgray}{Acc$\uparrow$} & \textcolor{EMgray}{Acc$\uparrow$} \\
\midrule
T=2.08s &  11.87 &  \textcolor{red}{93.30} &  93.28 &  78.46 &  61.75 &\textcolor{EMgray}{73.57}&\textcolor{EMgray}{\textbf{53.76}}\\
T=3.04s &   9.81 &  \textcolor{red}{94.99} &  95.15 &  79.18 &  63.39 &\textcolor{EMgray}{74.36}&\textcolor{EMgray}{52.14}\\
T=4.00s &   8.50 &  \textcolor{red}{95.04} &  94.83 &\textbf{81.24}&  63.81 &\textcolor{EMgray}{74.11}&\textcolor{EMgray}{52.23}\\
T=5.12s &   8.10 &  \textcolor{red}{95.24} &  94.70 &  78.73 &\textbf{65.47}&\textcolor{EMgray}{\textbf{74.69}}&\textcolor{EMgray}{51.66}\\
T=6.08s &\textbf{7.74}&\textcolor{red}{\textbf{96.41}}&\textbf{95.50}&  80.48 &  64.06 &\textcolor{EMgray}{72.42}&\textcolor{EMgray}{50.64}\\
\bottomrule\\
\end{tabular}
}
\vspace{-10pt}
\caption{Original table}
\label{original:tab:exp-input-dur}
\end{subtable}
\hfill
\begin{subtable}[t]{0.49\textwidth}
\centering
\resizebox{\textwidth}{!}{%
\begin{tabular}{llllll|ll}
\toprule
 &    PR &     KS &     IC &     SID &    ER & \textcolor{EMgray}{ENV}  & \textcolor{EMgray}{MUS} \\
\vspace{-1pt} Input duration $T$   &  PER$\downarrow$ & Acc$\uparrow$& Acc$\uparrow$& Acc$\uparrow$& Acc$\uparrow$& \textcolor{EMgray}{Acc$\uparrow$} & \textcolor{EMgray}{Acc$\uparrow$} \\
\midrule
T=2.08s &  11.87 &  \textcolor{red}{96.99} &  93.28 &  78.46 &  61.75 &\textcolor{EMgray}{73.57}&\textcolor{EMgray}{\textbf{53.76}}\\
T=3.04s &   9.81 &  \textcolor{red}{97.09} &  95.15 &  79.18 &  63.39 &\textcolor{EMgray}{74.36}&\textcolor{EMgray}{52.14}\\
T=4.00s &   8.50 &\textcolor{red}{\textbf{97.34}}&  94.83 &\textbf{81.24}&  63.81 &\textcolor{EMgray}{74.11}&\textcolor{EMgray}{52.23}\\
T=5.12s &   8.10 &  \textcolor{red}{97.17} &  94.70 &  78.73 &\textbf{65.47}&\textcolor{EMgray}{\textbf{74.69}}&\textcolor{EMgray}{51.66}\\
T=6.08s &\textbf{7.74}&  \textcolor{red}{97.17} &\textbf{95.50}&  80.48 &  64.06 &\textcolor{EMgray}{72.42}&\textcolor{EMgray}{50.64}\\
\bottomrule\\
\end{tabular}
}
\vspace{-10pt}
\caption{Corrected table}
\label{ccorrected:tab:exp-input-dur}
\hfill
\end{subtable}
\vspace{-5pt}
\end{table}

\begin{table}[tbp!]
\caption{Corrections of Table 5: M2D-S pre-training task ablations.
\footnotesize{(input duration $T=2.08s$, and patch size $80\times 4$)}}
\label{corr:tab:exp-tasks}
\vspace{-5pt}
\begin{subtable}[t]{0.49\textwidth}
\centering
\resizebox{\textwidth}{!}{%
\begin{tabular}{lllllllll|ll}
\toprule
&\multicolumn{2}{c}{Offline} & Online &    PR &     KS &     IC &     SID &    ER & \textcolor{EMgray}{ENV}  & \textcolor{EMgray}{MUS} \\
 \cmidrule(lr){2-3} \cmidrule(lr){4-4}
&\vspace{-1pt} denoise & distill. & M2D &  PER$\downarrow$ & Acc$\uparrow$& Acc$\uparrow$& Acc$\uparrow$& Acc$\uparrow$& \textcolor{EMgray}{Acc$\uparrow$} & \textcolor{EMgray}{Acc$\uparrow$} \\
\midrule
(a)&&&\checkmark                     &  11.87 &  \textcolor{red}{93.30} &  93.28 &\textbf{78.46}&  61.75 &\textcolor{EMgray}{\textbf{73.57}}&\textcolor{EMgray}{\textbf{53.76}}\\
(b)&&\checkmark&           &   8.10 &  \textcolor{red}{94.33} &  94.81 &  65.51 &  61.27 &\textcolor{EMgray}{56.96}&\textcolor{EMgray}{44.16}\\
(c)&\checkmark&\checkmark& &   7.80 &  \textcolor{red}{93.88} &  93.38 &  70.03 &  61.58 &\textcolor{EMgray}{42.42}&\textcolor{EMgray}{35.36}\\
(d)&&\checkmark&\checkmark                      &   8.78 &  \textcolor{red}{90.90} &  91.35 &  62.68 &  58.63 &\textcolor{EMgray}{57.15}&\textcolor{EMgray}{43.75}\\
(e)&\checkmark&\checkmark&\checkmark                            &\textbf{7.02}&\textcolor{red}{\textbf{97.36}}&\textbf{97.10}&  78.12 &\textbf{64.09}&\textcolor{EMgray}{55.49}&\textcolor{EMgray}{41.51}\\
\bottomrule\\
\end{tabular}
}
\vspace{-10pt}
\caption{Original table}
\label{original:tab:exp-tasks}
\end{subtable}
\hfill
\begin{subtable}[t]{0.49\textwidth}
\centering
\resizebox{\textwidth}{!}{%
\begin{tabular}{lllllllll|ll}
\toprule
&\multicolumn{2}{c}{Offline} & Online &    PR &     KS &     IC &     SID &    ER & \textcolor{EMgray}{ENV}  & \textcolor{EMgray}{MUS} \\
 \cmidrule(lr){2-3} \cmidrule(lr){4-4}
&\vspace{-1pt} denoise & distill. & M2D &  PER$\downarrow$ & Acc$\uparrow$& Acc$\uparrow$& Acc$\uparrow$& Acc$\uparrow$& \textcolor{EMgray}{Acc$\uparrow$} & \textcolor{EMgray}{Acc$\uparrow$} \\
\midrule
(a)&&&\checkmark        &  11.87 &\textcolor{red}{\textbf{96.99}}&  93.28 &\textbf{78.46}&  61.75 &\textcolor{EMgray}{\textbf{73.57}}&\textcolor{EMgray}{\textbf{53.76}}\\
(b)&&\checkmark&           &   8.10 &  \textcolor{red}{96.05} &  94.81 &  65.51 &  61.27 &\textcolor{EMgray}{56.96}&\textcolor{EMgray}{44.16}\\
(c)&\checkmark&\checkmark& &   7.80 &  \textcolor{red}{94.66} &  93.38 &  70.03 &  61.58 &\textcolor{EMgray}{42.42}&\textcolor{EMgray}{35.36}\\
(d)&&\checkmark&\checkmark                      &   8.78 &  \textcolor{red}{95.72} &  91.35 &  62.68 &  60.85 &\textcolor{EMgray}{57.15}&\textcolor{EMgray}{43.75}\\
(e)&\checkmark&\checkmark&\checkmark      &\textbf{7.02}&  \textcolor{red}{95.48} &\textbf{97.10}&  78.12 &\textbf{64.09}&\textcolor{EMgray}{55.49}&\textcolor{EMgray}{41.51}\\
\bottomrule\\
\end{tabular}
}
\vspace{-10pt}
\caption{Corrected table}
\label{ccorrected:tab:exp-tasks}
\hfill
\end{subtable}
\vspace{-5pt}
\end{table}

\begin{table}[tbp!]
\caption{Corrections of Table 6: Comparison with SOTA speech models. 
\scriptsize{($\lambda_\text{off}=0.5$, $\lambda_\text{m2d}=1$, $\alpha=0.2$, and patch size $80\times 2$)}}
\label{corr:tab:results-sota}
\vspace{-5pt}
\begin{subtable}[t]{0.49\textwidth}
\centering
\resizebox{\columnwidth}{!}{%
\begin{tabular}{lllllll|ll}
\toprule
 & &    PR &     KS &     IC &     SID &    ER & \textcolor{EMgray}{ENV}  & \textcolor{EMgray}{MUS} \\
\vspace{-1pt} Model     & Dataset &  PER$\downarrow$ & Acc$\uparrow$& Acc$\uparrow$& Acc$\uparrow$& Acc$\uparrow$& \textcolor{EMgray}{Acc$\uparrow$} & \textcolor{EMgray}{Acc$\uparrow$} \\
\midrule
wav2vec2.0 Base \cite{baevski2020wav2vec2}$^{\dagger}$ & \footnotesize{LS-960} & 5.74 & 96.23 & 92.35 & 75.18 & 63.43 & \textcolor{EMgray}{\textit{37.66}} &  \textcolor{EMgray}{\textit{32.02}} \\
HuBERT Base \cite{Hsu2021HuBERT}$^{\dagger}$ & \footnotesize{LS-960} & 5.41 & 96.30 & 98.34 & 81.42 & 64.92 & \textcolor{EMgray}{\textit{62.76}} & \textcolor{EMgray}{\textit{46.26}} \\
WavLM Base \cite{Chen2022WavLM}$^{\dagger}$ & \scriptsize{LS-960+DNS} & \textbf{4.84} & 96.79 & \textbf{98.63} & \textbf{84.51} & 65.94 & \textcolor{EMgray}{\textit{54.45}} & \textcolor{EMgray}{\textit{40.98}} \\
\midrule
\multicolumn{6}{l}{\textit{(Proposed using ViT \textit{Base})}} &&& \\
M2D-S T=4.0s  & \footnotesize{LS-960+AS} &   5.72 &  \textcolor{red}{97.80} &  97.80 &  81.97 &  \textbf{66.36} &\textcolor{EMgray}{53.22}&\textcolor{EMgray}{41.71}\\
M2D-S T=5.12s   & \footnotesize{LS-960+AS} &   5.64 &  \textcolor{red}{97.76} &  97.65 &  80.69 &  65.35 &\textcolor{EMgray}{57.34}&\textcolor{EMgray}{43.23}\\

M2D-S T=6.08s & \footnotesize{LS-960+AS} &   5.33 & \textcolor{red}{\textbf{97.81}} &  97.63 &  81.74 &    66.13 &\textcolor{EMgray}{54.77}&\textcolor{EMgray}{43.75}\\

\multicolumn{6}{l}{\textit{(Conventional using ViT \textit{Base})}} &&& \\
M2D ratio=0.6 \cite{niizumi2022M2D}$^{\natural}$ & \scriptsize{AS} &  78.30 &  \textcolor{red}{80.23} &  76.77 &  80.68 &  61.17 &  \textcolor{EMgray}{\textbf{88.63}}&\textcolor{EMgray}{\textbf{66.56}}\\
\midrule
\multicolumn{6}{l}{\textit{(Reference \textit{Large} models)}} &&& \\
\textcolor{gray}{wav2vec\small{2.0 Large \cite{baevski2020wav2vec2}}$^{\dagger}$} & \textcolor{gray}{\footnotesize{LL-60k}} & \textcolor{gray}{4.75} & \textcolor{gray}{96.66} & \textcolor{gray}{95.28} & \textcolor{gray}{86.14} & \textcolor{gray}{65.64} & \textcolor{gray}{\textit{60.82}} &  \textcolor{gray}{\textit{42.75}} \\
\textcolor{gray}{HuBERT Large \cite{Hsu2021HuBERT}$^{\dagger}$} & \textcolor{gray}{\footnotesize{LL-60k}} & \textcolor{gray}{3.53} & \textcolor{gray}{95.29} & \textcolor{gray}{98.76} & \textcolor{gray}{90.33} & \textcolor{gray}{67.62} & \textcolor{gray}{\textit{59.51}} &  \textcolor{gray}{\textit{44.35}} \\
\textcolor{gray}{WavLM Large \cite{Chen2022WavLM}$^{\dagger}$} & \textcolor{gray}{\footnotesize{Mix-94k}} & \textcolor{gray}{\textbf{3.06}} & \textcolor{gray}{97.86} & \textcolor{gray}{\textbf{99.31}} & \textcolor{gray}{\textbf{95.49}} & \textcolor{gray}{\textbf{70.62}} & \textcolor{gray}{\textit{69.32}} &  \textcolor{gray}{\textit{50.56}} \\
\bottomrule
\addlinespace[0.1cm]
\multicolumn{9}{l}{$^{\dagger}$
ENV and MUS results were obtained using publicly available pre-trained models.}\\
\multicolumn{9}{l}{$^{\natural}$
The original M2D takes input with $T=6.08s$ and uses a patch size of $16\times 16$.}\\
\end{tabular}
}
\caption{Original table}
\label{original:tab:results-sota}
\end{subtable}
\hfill
\begin{subtable}[t]{0.49\textwidth}
\centering
\resizebox{\textwidth}{!}{%
\begin{tabular}{lllllll|ll}
\toprule
 & &    PR &     KS &     IC &     SID &    ER & \textcolor{EMgray}{ENV}  & \textcolor{EMgray}{MUS} \\
\vspace{-1pt} Model     & Dataset &  PER$\downarrow$ & Acc$\uparrow$& Acc$\uparrow$& Acc$\uparrow$& Acc$\uparrow$& \textcolor{EMgray}{Acc$\uparrow$} & \textcolor{EMgray}{Acc$\uparrow$} \\
\midrule
wav2vec2.0 Base \cite{baevski2020wav2vec2}$^{\dagger}$ & \footnotesize{LS-960} & 5.74 & 96.23 & 92.35 & 75.18 & 63.43 & \textcolor{EMgray}{\textit{37.66}} &  \textcolor{EMgray}{\textit{32.02}} \\
HuBERT Base \cite{Hsu2021HuBERT}$^{\dagger}$ & \footnotesize{LS-960} & 5.41 & 96.30 & 98.34 & 81.42 & 64.92 & \textcolor{EMgray}{\textit{62.76}} & \textcolor{EMgray}{\textit{46.26}} \\
WavLM Base \cite{Chen2022WavLM}$^{\dagger}$ & \scriptsize{LS-960+DNS} & \textbf{4.84} & 96.79 & \textbf{98.63} & \textbf{84.51} & 65.94 & \textcolor{EMgray}{\textit{54.45}} & \textcolor{EMgray}{\textit{40.98}} \\
\midrule
\multicolumn{6}{l}{\textit{(Proposed using ViT \textit{Base})}} &&& \\
M2D-S T=4.0s  & \footnotesize{LS-960+AS} &   5.72 &  \textcolor{red}{96.47} &  97.80 &  81.97 &  \textbf{66.36} &\textcolor{EMgray}{53.22}&\textcolor{EMgray}{41.71}\\
M2D-S T=5.12s   & \footnotesize{LS-960+AS} &   5.64 &  \textcolor{red}{\textbf{96.87}} &  97.65 &  80.69 &  65.35 &\textcolor{EMgray}{57.34}&\textcolor{EMgray}{43.23}\\

M2D-S T=6.08s & \footnotesize{LS-960+AS} &   5.33 & \textcolor{red}{96.80} &  97.63 &  81.74 &    66.13 &\textcolor{EMgray}{54.77}&\textcolor{EMgray}{43.75}\\

\multicolumn{6}{l}{\textit{(Conventional using ViT \textit{Base})}} &&& \\
M2D ratio=0.6 \cite{niizumi2022M2D}$^{\natural}$ & \scriptsize{AS} &  78.30 &  \textcolor{red}{95.65} &  76.77 &  80.68 &  61.17 &  \textcolor{EMgray}{\textbf{88.63}}&\textcolor{EMgray}{\textbf{66.56}}\\
\midrule
\multicolumn{6}{l}{\textit{(Reference \textit{Large} models)}} &&& \\
\textcolor{gray}{wav2vec\small{2.0 Large \cite{baevski2020wav2vec2}}$^{\dagger}$} & \textcolor{gray}{\footnotesize{LL-60k}} & \textcolor{gray}{4.75} & \textcolor{gray}{96.66} & \textcolor{gray}{95.28} & \textcolor{gray}{86.14} & \textcolor{gray}{65.64} & \textcolor{gray}{\textit{60.82}} &  \textcolor{gray}{\textit{42.75}} \\
\textcolor{gray}{HuBERT Large \cite{Hsu2021HuBERT}$^{\dagger}$} & \textcolor{gray}{\footnotesize{LL-60k}} & \textcolor{gray}{3.53} & \textcolor{gray}{95.29} & \textcolor{gray}{98.76} & \textcolor{gray}{90.33} & \textcolor{gray}{67.62} & \textcolor{gray}{\textit{59.51}} &  \textcolor{gray}{\textit{44.35}} \\
\textcolor{gray}{WavLM Large \cite{Chen2022WavLM}$^{\dagger}$} & \textcolor{gray}{\footnotesize{Mix-94k}} & \textcolor{gray}{\textbf{3.06}} & \textcolor{gray}{97.86} & \textcolor{gray}{\textbf{99.31}} & \textcolor{gray}{\textbf{95.49}} & \textcolor{gray}{\textbf{70.62}} & \textcolor{gray}{\textit{69.32}} &  \textcolor{gray}{\textit{50.56}} \\
\bottomrule
\addlinespace[0.1cm]
\multicolumn{9}{l}{$^{\dagger}$
{ENV and MUS results were obtained using publicly available pre-trained models.}}\\
\multicolumn{9}{l}{$^{\natural}$
{The original M2D takes input with $T=6.08s$ and uses a patch size of $16\times 16$.}}\\
\end{tabular}
}
\caption{Corrected table}
\label{ccorrected:tab:results-sota}
\end{subtable}
\vspace{-10pt}
\end{table}

\section{Comparing with recent speech models}

Table \ref{tab:results-sota-more} shows an updated version of Table 6 with SSAST~\cite{gong2022ssast} and Whisper~\cite{radford2022whisper, chemudupati2023transferwhisper} results based on suggestions received from the reviewer's comments.
The KS results in Table \ref{tab:results-sota-more} have been corrected.

\begin{table*}[tb!]
\vspace{-10pt}
\caption{Comparison with more SOTA speech models. \\
\footnotesize{($\lambda_\text{off}=0.5$, $\lambda_\text{m2d}=1$, $\alpha=0.2$, and patch size $80\times 2$)}}
\vspace{-5pt}
\label{tab:results-sota-more}
\centering
\begin{tabular}{lllllll|ll}
\toprule
 & &    PR &     KS &     IC &     SID &    ER & \textcolor{EMgray}{ENV}  & \textcolor{EMgray}{MUS} \\
\vspace{-1pt} Model     & Dataset &  PER$\downarrow$ & Acc$\uparrow$& Acc$\uparrow$& Acc$\uparrow$& Acc$\uparrow$& \textcolor{EMgray}{Acc$\uparrow$} & \textcolor{EMgray}{Acc$\uparrow$} \\
\midrule
wav2vec2.0 Base \cite{baevski2020wav2vec2}$^{\dagger}$ & \footnotesize{LS-960} & 5.74 & 96.23 & 92.35 & 75.18 & 63.43 & \textcolor{EMgray}{\textit{37.66}} &  \textcolor{EMgray}{\textit{32.02}} \\
HuBERT Base \cite{Hsu2021HuBERT}$^{\dagger}$ & \footnotesize{LS-960} & 5.41 & 96.30 & 98.34 & 81.42 & 64.92 & \textcolor{EMgray}{\textit{62.76}} & \textcolor{EMgray}{\textit{46.26}} \\
WavLM Base~\cite{Chen2022WavLM}$^{\dagger}$ & \scriptsize{LS-960+DNS} & \textbf{4.84} & 96.79 & \textbf{98.63} & \textbf{84.51} & 65.94 & \textcolor{EMgray}{\textit{54.45}} & \textcolor{EMgray}{\textit{40.98}} \\
\midrule
\multicolumn{6}{l}{\textit{(Proposed using ViT \textit{Base})}} &&& \\
M2D-S T=4.0s  & \footnotesize{LS-960+AS} &   5.72 &  96.47 &  97.80 &  81.97 &  \textbf{66.36} &\textcolor{EMgray}{53.22}&\textcolor{EMgray}{41.71}\\
M2D-S T=5.12s   & \footnotesize{LS-960+AS} &   5.64 &  \textbf{96.87} &  97.65 &  80.69 &  65.35 &\textcolor{EMgray}{57.34}&\textcolor{EMgray}{43.23}\\

M2D-S T=6.08s & \footnotesize{LS-960+AS} &   5.33 & {96.80} &  97.63 &  81.74 &    66.13 &\textcolor{EMgray}{54.77}&\textcolor{EMgray}{43.75}\\

\multicolumn{6}{l}{\textit{(Conventional using ViT \textit{Base} or similar models)}} &&& \\
M2D ratio=0.6, T=6.08s, patch size $16\times 16$ & \footnotesize{AS} &  78.30 &  95.65 &  76.77 &  80.68 &  61.17 &  \textcolor{EMgray}{\textbf{88.63}}&\textcolor{EMgray}{\textbf{66.56}}\\
SSAST-Frame~\cite{gong2022ssast} & \footnotesize{LS-960 $\cup$ AS}$~^{\natural}$ & - & 96.7 & - & 80.8 & 60.5 & - & - \\
SSAST-Patch~\cite{gong2022ssast} & \footnotesize{LS-960 $\cup$ AS}$~^{\natural}$ & - & 94.8 & - & 57.1 & 56.8 & - & - \\

\midrule
\multicolumn{6}{l}{\textit{(Reference non-SSL models)}} &&& \\
\textcolor{gray}{Whisper Base~\cite{radford2022whisper, chemudupati2023transferwhisper}} & \textcolor{gray}{\footnotesize{680K hr (labeled)}} & \textcolor{gray}{-} & \textcolor{gray}{97.63} & \textcolor{gray}{95.17} & \textcolor{gray}{64.96} & \textcolor{gray}{7.57} & \textcolor{gray}{-} & \textcolor{gray}{-}  \\
\textcolor{gray}{Whisper Base (fine-tuned)~\cite{radford2022whisper, chemudupati2023transferwhisper}} & \textcolor{gray}{\footnotesize{680K hr (labeled)}} & \textcolor{gray}{-} & \textcolor{gray}{95.58} & \textcolor{gray}{\textbf{99.45}} & \textcolor{gray}{68.87} & \textcolor{gray}{\textbf{84.85}} & \textcolor{gray}{-} & \textcolor{gray}{-}  \\

\multicolumn{6}{l}{\textit{(Reference \textit{Large} models)}} &&& \\
\textcolor{gray}{wav2vec\small{2.0 Large \cite{baevski2020wav2vec2}}$^{\dagger}$} & \textcolor{gray}{\footnotesize{LL-60k}} & \textcolor{gray}{4.75} & \textcolor{gray}{96.66} & \textcolor{gray}{95.28} & \textcolor{gray}{86.14} & \textcolor{gray}{65.64} & \textcolor{gray}{\textit{60.82}} &  \textcolor{gray}{\textit{42.75}} \\
\textcolor{gray}{HuBERT Large \cite{Hsu2021HuBERT}$^{\dagger}$} & \textcolor{gray}{\footnotesize{LL-60k}} & \textcolor{gray}{3.53} & \textcolor{gray}{95.29} & \textcolor{gray}{98.76} & \textcolor{gray}{90.33} & \textcolor{gray}{67.62} & \textcolor{gray}{\textit{59.51}} &  \textcolor{gray}{\textit{44.35}} \\
\textcolor{gray}{WavLM Large \cite{Chen2022WavLM}$^{\dagger}$} & \textcolor{gray}{\footnotesize{Mix-94k}} & \textcolor{gray}{\textbf{3.06}} & \textcolor{gray}{97.86} & \textcolor{gray}{{99.31}} & \textcolor{gray}{\textbf{95.49}} & \textcolor{gray}{{70.62}} & \textcolor{gray}{\textit{69.32}} &  \textcolor{gray}{\textit{50.56}} \\
\bottomrule
\addlinespace[0.1cm]
\multicolumn{9}{l}{$^{\dagger}$
ENV and MUS results were obtained using publicly available pre-trained models.}\\
\multicolumn{9}{l}{$^{\natural}$ The LS-960 and the AS samples were used without mixing one as BG noise into others, unlike ours.}\\
\end{tabular}
\vspace{-15pt}
\end{table*}

\end{document}